\documentclass[12pt,prl,aps,reprint,superscriptaddress,twocolumn,notitlepage]{revtex4-2}

\usepackage{amssymb,amsmath}
\usepackage{graphicx}
\usepackage{xcolor} 
\usepackage[normalem]{ulem}
\usepackage{siunitx}
\usepackage[english]{babel}

\setcitestyle{numbers,square,sort&compress}

\usepackage{caption}

\usepackage[hidelinks]{hyperref}

\newcommand{\beginsupplement}{%
  \clearpage
  \onecolumngrid
  \setcounter{section}{0}%
  \setcounter{subsection}{0}%
  \setcounter{figure}{0}%
  \setcounter{table}{0}%
  \setcounter{equation}{0}%
  \renewcommand{\thesection}{S\arabic{section}}%
  \renewcommand{\thesubsection}{\thesection.\arabic{subsection}}%
  \renewcommand{\thefigure}{S\arabic{figure}}%
  \renewcommand{\thetable}{S\arabic{table}}%
  \renewcommand{\theequation}{S\arabic{equation}}%
  \captionsetup[figure]{name=Figure}%
  \captionsetup[table]{name=Table}%
}

\begin{document}

\newcommand{\e}{{\rm e}}
\newcommand{\norm}[1]{\left\lVert#1\right\rVert}
\newcommand{\rmi}{{\rm i}}
\renewcommand{\Im}{\mathop\mathrm{Im}\nolimits}
\renewcommand{\Re}{\mathop\mathrm{Re}\nolimits}
\newcommand{\red}[1]{{\color{red}#1}}
\newcommand{\cyan}[1]{{\color{cyan}#1}}
\newcommand{\blue}[1]{{\color{blue}#1}}
\newcommand{\rot}[1]{\text{rot}\, #1}
\newcommand{\df}[2]{\frac{\partial #1}{\partial #2}}
\newcommand{\bas}[1]{\hat{{\bf #1}}}
\renewcommand{\cite}[1]{[\onlinecite{#1}]}

\newcommand{\ket}[1]{\left|#1\right>}      
\newcommand{\bra}[1]{\left<#1\right|}
\newcommand{\eps}{\varepsilon}      
\newcommand{\om}{\omega}      
\newcommand{\kap}{\varkappa}      
\newcommand{\VB}[1]{\mathbf{#1}} 

\renewcommand{\thefootnote}{\roman{footnote}}

\newcommand{\affilITMO}{Department of Physics and Engineering, ITMO University,  St.-Petersburg 197101, Russia}


\title{Emergence of transverse optical spin in a subwavelength grating ring resonator}


\author{Nikita Iukhtanov}
 \affiliation{Department of Physics and Engineering, ITMO University, St.-Petersburg 197101, Russia}

\author{Chao Sun}
\email{chaosun@hrbeu.edu.cn}
\affiliation{College of Physics and Optoelectronic Engineering, Harbin Engineering University, Harbin 150001, China}

\author{Georgiy Kurganov}
 \affiliation{Department of Physics and Engineering, ITMO University, St.-Petersburg 197101, Russia}
 
\author{Dmitry Zhirihin}
 \affiliation{Department of Physics and Engineering, ITMO University, St.-Petersburg 197101, Russia}

\author{Andrey Bogdanov}
\affiliation{College of Physics and Optoelectronic Engineering, Harbin Engineering University, Harbin 150001, China}

 \altaffiliation{Department of Physics and Engineering, ITMO University, St.-Petersburg 197101, Russia}

\author{Roman Savelev}
\email{r.savelev@metalab.ifmo.ru}
 \affiliation{Department of Physics and Engineering, ITMO University, St.-Petersburg 197101, Russia}

\begin{abstract}
The local polarization of the electromagnetic field plays a crucial role in the interaction of light with spin- and valley-polarized quantum sources. Unlike free-space electromagnetic waves, whose polarization degeneracy enables flexible polarization manipulation, planar integrated optical structures lack such degree of freedom owing to intrinsic structural anisotropy. Here, we propose a planar optical ring resonator based on a subwavelength grating waveguide that supports two quasi-degenerate modes. We demonstrate that coupling of these modes in the ring resonator leads to the formation of the resonances with a predominant direction of electric-field rotation in the vicinity of the resonator, resulting in the non-zero transverse optical spin. The average degree of circular polarization in the proposed structures reaches values of up to 70\%. The theoretical predictions are corroborated by experimental validation in the microwave spectral range. Our findings suggest a viable route toward realization of on-chip optical spintronic and valleytronic interfaces.
\end{abstract}

\maketitle

\section{Introduction}

Recently, there has been growing interest in nanophotonic systems incorporating spin- or valley-polarized excitons in semiconductors~\cite{LodahlNature2017,ChenAOM2020,XiaoLPR2021,AziminpjNP2025,ForeroPRXQ2025}. Such systems have already enabled demonstrations of unidirectional spin-to-path coupling, single-photon sources, and quantum gates operating at the single-photon level, among other functionalities~\cite{ForeroPRXQ2025}. Selective addressing of the polarization degree of freedom of the excitons becomes possible with the development of efficient spin–photon interfaces, that is, optical structures capable of manipulating one of the key characteristics of these systems -- the degree of circular polarization (DoCP) of the local electric field, or equivalently, the electric component of the optical spin density~\cite{BliokhPR2015,PicardiOPT2018}. Among the various structures investigated for this purpose, integrated optical platforms -- such as different types of planar optical waveguides~\cite{SollnerNN2015,ColesNC2016,JavadiNN2018,GongSC2018,YangOE2019,GongNL2020,WooOE2021,XiaoAPL2021,ChenACSnano2021,ChenNN2022,MartinPRR2024,GermanisOptica2025} and resonators~\cite{CanoACS2019,BarikPRB2020,MehrabadOptica2020,MehrabadAPL2020,GuddalaNC2021,HallettACS2022,ShiAPL2023,RaoNL2025,MartinOPT2025,LiLPR2026} combined with quantum dots and quasi-two-dimensional materials -- have attracted particular attention. Their main advantage lies in the maturity of integrated photonic circuit platforms, which enables the practical realization of application-oriented photonic components and devices for data processing, sensing, and quantum technologies.

State-of-the-art integrated spin-photonic elements based on photonic crystal waveguides and resonators with embedded quantum dots~\cite{SollnerNN2015,MartinPRR2024,RaoNL2025,MartinOPT2025} achieve near-perfect interaction unidirectionality and a fraction of photons emitted into a given mode approaching unity~\cite{MahmoodianOME2017,SiampourQI2023,MartinPRR2024,MartinOPT2025}. However, the asymmetry of interaction between the optical mode and right and left circularly-polarized (CP) transitions in a quantum source typically exhibits a strong dependence on the position of the source within the waveguide or resonator. This behavior arises from the pronounced spatial variation of the DoCP inherent to integrated photonic structures due to their structural anisotropy. Consequently, the implementation of unidirectional light–matter interaction with single quantum dots requires extremely precise positioning within the optical structure, often with nanometer-scale accuracy~\cite{ChuAQT2020,XiaoAPL2021,MartinPRR2024,MartinOPT2025,RaoNL2025}. Importantly, for distributed emitters, such as excitons in quasi-two-dimensional materials, the challenge becomes even greater. Owing to symmetry, the DoCP averaged over the area above a typical single-mode waveguide or ring resonator vanishes. As a result, achieving asymmetric light–matter interaction necessitates carefully engineered asymmetric waveguides, precise placement of monolayer flakes, or accurate control of the excitation spot position~\cite{GongSC2018,YangOE2019,SunNP2019,GongNL2020,WooOE2021,ChenACSnano2021,GuddalaNC2021}, making valleytronic devices technologically demanding.

\begin{figure*}[t]
\centering
    \includegraphics[width=0.8\textwidth]{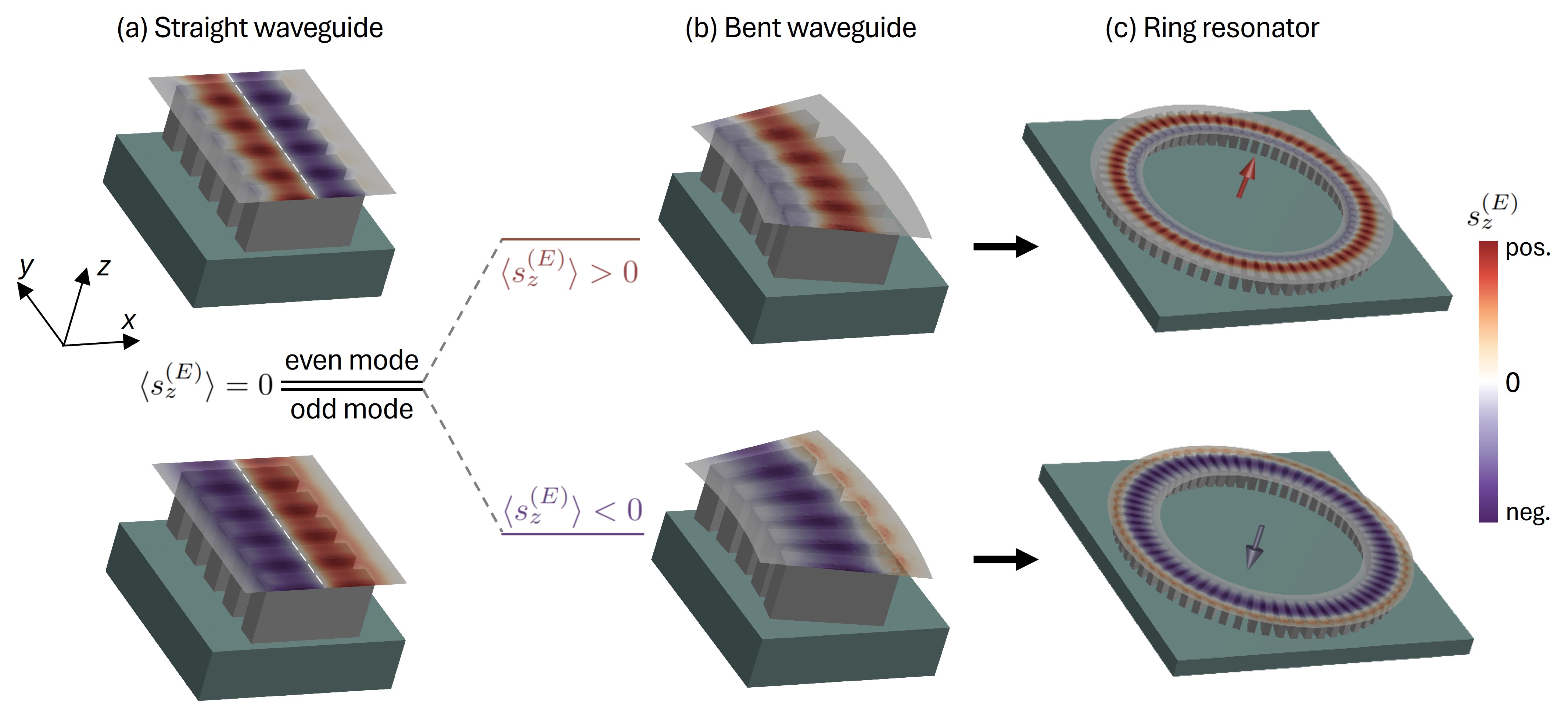}
\caption{(a) Two quasi-degenerate modes of a SWG waveguide exhibit antisymmetric distributions of the
z-component of the electric contribution to the optical spin density, $s_z^{(E)}$. The unit-cell average $\langle s_z^{(E)} \rangle$ vanishes for both modes. (b) Bending of the waveguide breaks the mirror symmetry and induces hybridization of the two modes, resulting in a finite unit-cell-averaged spin density, $\langle s_z^{(E)} \rangle \ne 0$. (c) Hybridized counter-clockwise propagating resonant modes in a ring resonator exhibit a quasi-uniform sign of $s_z^{(E)}$, yielding non-zero $\langle s_z^{(E)} \rangle$ with opposite signs for the two resonances, indicated by arrows in the centres of the rings.}
	\label{fig:scheme}
\end{figure*}

In our work, we address this problem by proposing a special design of an integrated optical ring resonator that allows achieving predominant polarization handedness of the local electric field in the vicinity of the structure. The periodic ring resonator is based on a subwavelength grating (SWG) waveguide ~\cite{HalirLPR2015,ChebenNature2018,ChebenAOP2023}, shown schematically in Fig.~\ref{fig:scheme}(a). The parameters of the SWG waveguide are tuned so that it supports two accidentally degenerate modes, which serves as an additional degree of freedom that grants partial control over polarization of the local electric field. We investigate theoretically interaction of the waveguide modes when the waveguide is bent into a ring, revealing the formation of the ring resonances with predominant direction of electric field rotation in the plane of the resonator, Figs.~\ref{fig:scheme}(b,c). We show that in contrast to a conventional ring resonator based on the single-mode waveguide, in which the DoCP averaged over the waveguide cross-section is approximately zero due to symmetry of the mode, in the considered periodic ring resonator average DoCP can reach values up to tens of percent. We support our theoretical and numerical analysis with the experimental results obtained for the resonator composed of ceramic particles in the microwave spectral range.

\section{Integrated optical two-mode ring resonator}

It is known that the evanescent field of a guided mode is locally elliptically polarized due to a $\pi/2$ phase shift between its transverse and longitudinal components ~\cite{BliokhNP2015}. This leads to asymmetric light-matter coupling between guided waves and quantum sources that exhibit circularly polarized $\sigma^+$ and $\sigma^-$ dipole transition moments~\cite{VanMechelenOPT2016}. The resulting coupling asymmetry can be quantified by the degree of circular polarization (DoCP) of the electric field, denoted by $C$. For a field rotating in the $xy$ plane, $C$ is defined as:
\begin{equation}
\label{eq:DoCP}
C(\VB{r}) = \dfrac{|E_r(\VB{r})|^2 - |E_l(\VB{r})|^2}{|E_r(\VB{r})|^2 + |E_l(\VB{r})|^2} = \dfrac{\Im{(\VB{E(\VB{r})}^*\times\VB{E(\VB{r})})_z}}{|E_r(\VB{r})|^2 + |E_l(\VB{r})|^2},
\end{equation}
where $E_{r,l}(\VB{r}) = \VB{E}(\VB{r})\cdot (\hat{x} \mp i\hat{y})/\sqrt{2}$. Because the DoCP is normalized by $|E_r|^2 + |E_l|^2$, it captures only the relative asymmetry of the light–matter coupling, while information about the absolute coupling strength is discarded. Consequently, this metric is most informative for single-emitter configurations. 

For an ensemble of spatially distributed sources, two complementary measures of interaction asymmetry are more suitable.
The first is the numerator of~\eqref{eq:DoCP}, $|E_r(\VB{r})|^2 - |E_l(\VB{r})|^2$, which preserves the spatial distribution of the field intensity. This quantity is closely related to the optical spin density~\cite{PicardiOPT2018}:
\begin{multline}
\VB{s}(\VB{r}) = \VB{s}^{(E)}(\VB{r}) + \VB{s}^{(H)}(\VB{r}) =\\
=\dfrac{\eps_0 \eps}{4 \omega}\Im(\VB{E}^*(\VB{r})\times\VB{E}(\VB{r})) + \dfrac{\mu_0 \mu}{4 \omega}\Im(\VB{H}^*(\VB{r})\times\VB{H}(\VB{r})).
\end{multline}
Upon normalization by $\dfrac{\eps_0 \eps}{4 \omega}$, the electric contribution to the $z$-component of the spin density $\VB{s}$ can be written as
\begin{equation}
\label{eq:SE_norm}
\tilde{s}_z^{(E)}(\VB{r}) = |E_r(\VB{r})|^2 - |E_l(\VB{r})|^2,
\end{equation}
which we use throughout the paper to characterize the spatial distribution of the field polarization and spin density.

The second quantity we use to characterize the asymmetry of light–matter coupling for randomly distributed circular dipole sources ($\sigma^+$ or $\sigma^-$) is the average DoCP, defined as
\begin{equation}
\label{eq:DoCPav}
\langle C \rangle= \dfrac{\langle |E_r|^2 - |E_l|^2 \rangle}{\langle |E_r|^2 + |E_l|^2 \rangle},
\end{equation}
where $\langle...\rangle$ denotes averaging over the volume (or area) that contains the emitters. Note that, whereas the electric contribution to the spin density explicitly includes the permittivity of the surrounding medium, the DoCP depends only on the spatial distribution of the electric field. Consequently, $\langle C \rangle$ is a convenient metric for ensembles of emitters that are assumed to reside within the same material. Throughout the paper, we evaluate $\langle C \rangle$ by averaging over a plane parallel to the substrate, assuming coupling to excitons in a quasi-two-dimensional layer placed above the structure.

\begin{figure*}[t]
\centering
 \includegraphics[width=1\textwidth]{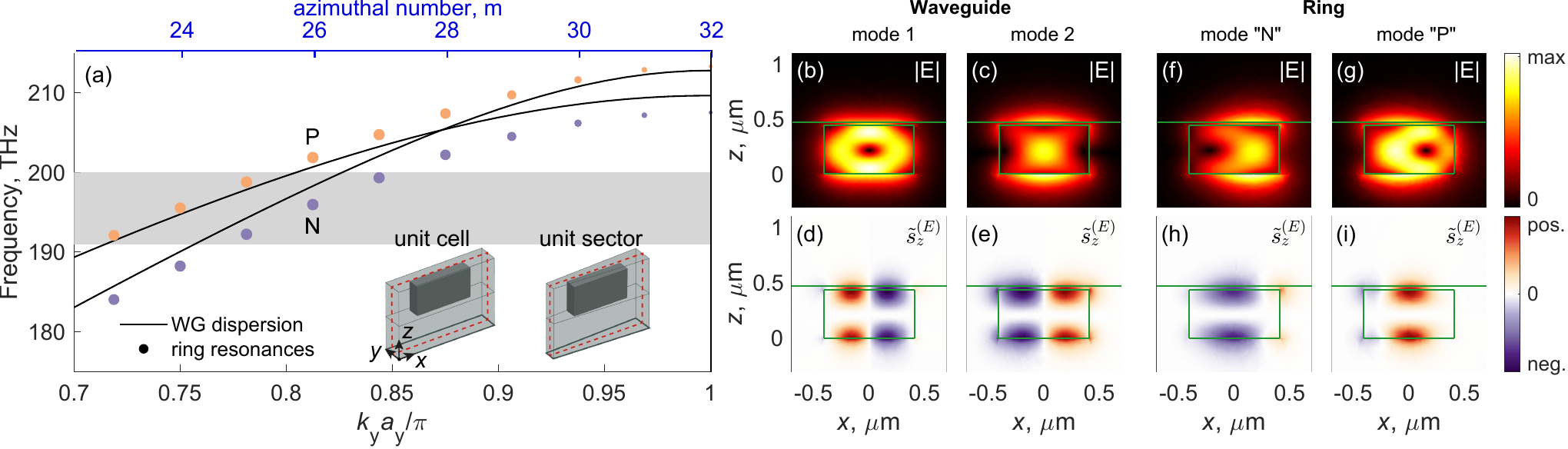}
	\caption{(a) Dispersion of the straight SWG waveguide (solid lines) and resonance frequencies of the ring resonator formed from this waveguide. Purple (orange) circles denote resonances with negative (positive) unit cell averaged $s_z^{(E)}$; the circle size encodes the magnitude of $s_z^{(E)}$ integrated over the unit cell. Structural parameters are given in the text. Insets show the waveguide unit cell and the corresponding unit sector of the ring. (b--e) Distributions of (b,c) the electric field and (d,e) the $z$-component of the electric contribution to the optical spin density, normalized to the maximum value of $|s_z^{(E)}|$ across both modes, for the two straight-waveguide modes near the degeneracy frequency ($\approx 202~\mathrm{THz}$). The fields are plotted in the $y=0$ plane, indicated by the red dashed contour in the inset of (a). (f--i) Same as (b--e), but for the two ring resonances with $m=26$.}
	\label{fig:optical_1}
\end{figure*}

The ring resonator is formed by SWG waveguide composed of silicon blocks (refractive index $n=3.5$) partially embedded in silicon dioxide ($\mathrm{SiO_2}$, refractive index $n_\mathrm{SiO_2}=1.45$, as shown schematically in the inset of Fig.~\ref{fig:optical_1}(a). The waveguide dispersion was engineered to achieve an accidental degeneracy between two modes, indicated by the black curves in Fig.~\ref{fig:optical_1}(a). The geometric parameters are as follows: $w_x=820$~nm, $w_z=440$~nm, $w_y=170$~nm, period $a_y=295$~nm, the $\mathrm{SiO_2}$ cladding above the silicon blocks has a thickness of 30~nm. For clarity, we show only the two modes of interest, which are quasi-even with respect to the horizontal symmetry plane $z=0$ passing through the center of the unit cell.

The field distributions of the waveguide modes at the mode-crossing (degeneracy) frequency are shown in Figs.~\ref{fig:optical_1}(b,c). Due to the mirror symmetry plane $x=0$, the corresponding distributions of $\tilde{s}_z^{(E)}$ are antisymmetric (odd under $x\to -x$). Consequently, both $\langle \tilde{s}_z^{(E)} \rangle$ and $\langle C \rangle$ vanish when averaged over the unit cell, as illustrated in Fig.~\ref{fig:scheme}(a) and Figs.~\ref{fig:optical_1}(d,e). In contrast, a linear superposition of these modes with a relative phase $\phi \neq 0,\pi$ yields a non-zero electric part of the total optical spin, quantified by the integral of $\tilde{s}_z^{(E)}$ over the unit cell of the waveguide. As elaborated in Sec.~S1 of the Supporting Information, the total spin is maximized at $\phi=\pm \pi/2$. This linear combination can be realized by introducing a geometrical perturbation that breaks the $x=0$ symmetry plane while preserving the $y=0$ symmetry~\cite{volkov2021unidirectional}. Physically, this corresponds to the waveguide being bent or, ultimately, closed into a ring resonator. Further details are provided in Sec.~S1 of the Supporting Information.

The resonance frequencies of the ring resonator composed of $N=64$ particles with radius $R_{\mathrm{ring}}=3$~$\mu$m ($\approx Na_y/(2\pi)$) are shown in Fig.~\ref{fig:optical_1}(a). Each eigenmode is assigned a wavenumber $k_y$ based on its azimuthal mode number $m$ via $k_y=m/R_{\mathrm{ring}}$. Owing to the discrete azimuthal periodicity, $m$ is bounded by $|m|\leq N/2$.
The total number of resonances is $2N$, corresponding to two mode types for both positive and negative $m$. As seen in Fig.~\ref{fig:optical_1}(a), the ring eigenmodes exhibit an avoided crossing near the degeneracy point, indicating mode coupling. Increasing the bending radius (equivalently, increasing the number of particles for a fixed waveguide period $a_y$) reduces the coupling strength, whereas the field profiles and hence the average spin remain nearly unchanged. 

The electric-field distributions in the resonator cross-section are shown in Figs.~\ref{fig:optical_1}(f,g). Both resonances correspond to the same positive azimuthal mode number $m=26$, as marked in Fig.~\ref{fig:optical_1}(a) by the labels ``N'' and ``P'' (negative and positive $\langle \tilde{s}_z^{(E)} \rangle$, respectively, averaged over the unit sector). As soon as the radius of the ring is large and the corresponding perturbation caused by bending is weak, the ring eigenmodes can be rather accurately approximated as a superposition of the coupled modes of a straight waveguide with a relative phase of $+\pi/2$ or $-\pi/2$. This superposition results in a non-zero unit cell averaged $\langle \tilde{s}_z^{(E)} \rangle$, as illustrated in Figs.~\ref{fig:optical_1}(h,i). In Fig.~\ref{fig:optical_1}(a), the magnitude of $\langle \tilde{s}_z^{(E)} \rangle$ is encoded by the circle size. The lower-frequency branch (purple circles) exhibits a negative $\langle \tilde{s}_z^{(E)} \rangle$ [Fig.~\ref{fig:optical_1}(h)], whereas the higher-frequency branch (orange circles) exhibits a positive $\langle \tilde{s}_z^{(E)} \rangle$ [Fig.~\ref{fig:optical_1}(i)]. For the $m=26$ resonance, the DoCP averaged over the plane immediately above the top oxide layer reaches $|\langle C \rangle|\approx 0.22$.

\begin{figure}[t]
\centering
\includegraphics[width=1\columnwidth]{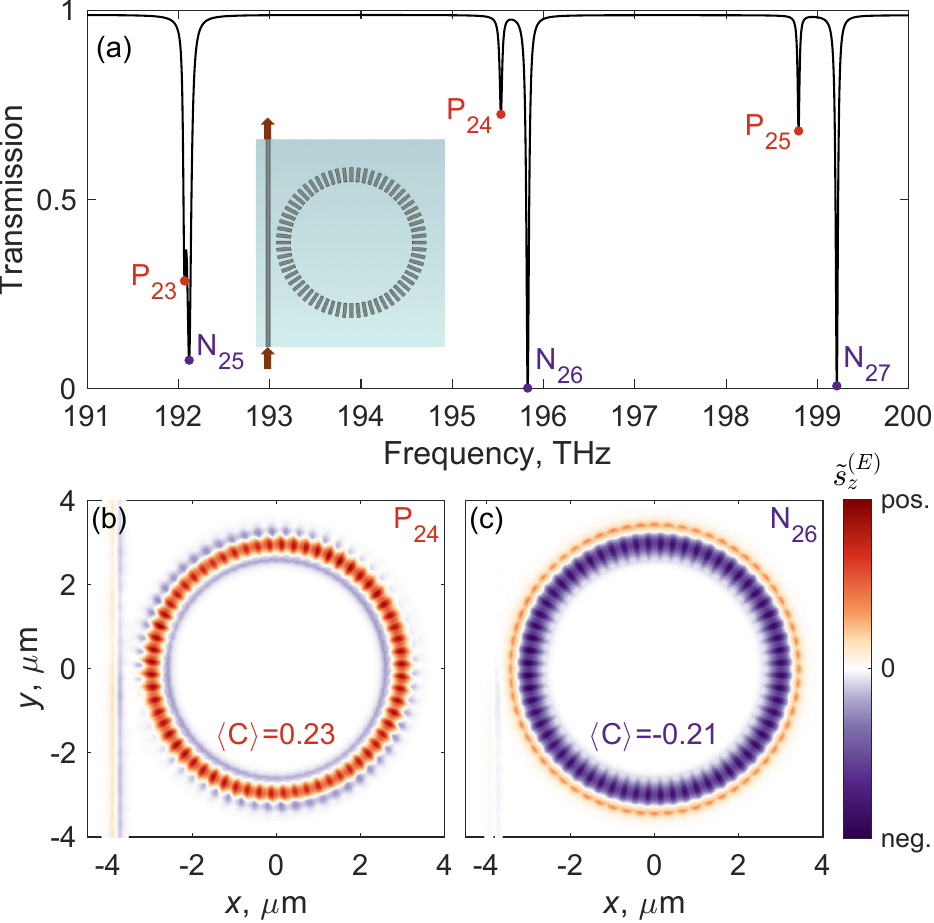}
	\caption{(a) Transmission spectra of the ring resonator excited by a bus waveguide, as shown in the inset.  Structural parameters are given in the text. The labels (P/N)$_m$ denote the ``P''/``N'' mode type with azimuthal number $m$. (b,c) Spatial distribution of the normalized optical spin density $\tilde{s}_z^{(E)}$ in a plane located $30~\mathrm{nm}$ above the ring surface, evaluated at the frequencies marked as P$_{24}$ and N$_{26}$ in panel~(a).}
	\label{fig:optical_2}
\end{figure}

\begin{figure*}[t]
\centering
\includegraphics[width=0.96\textwidth]{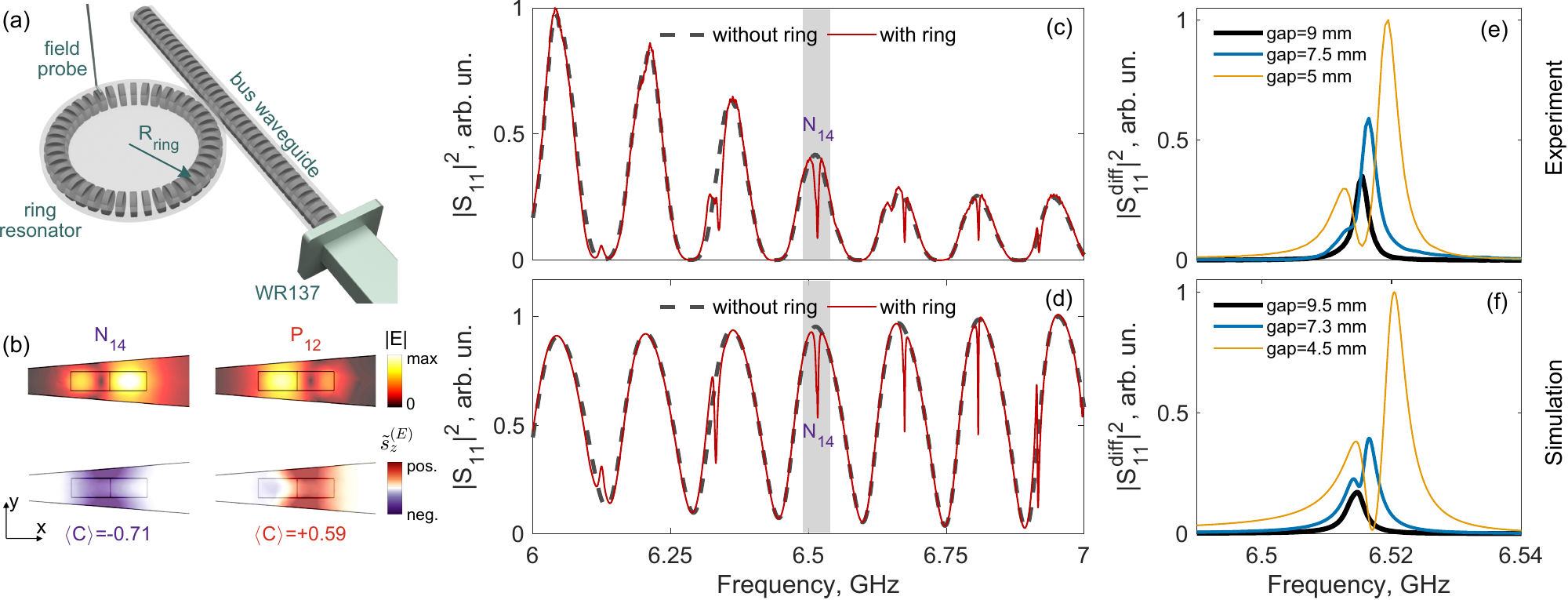}
	\caption{(a) Schematic of the experimental setup. (b) Spatial distributions of the electric-field amplitude (top panels) and the $z$-component of the optical spin density (bottom panels) for the $N_{14}$ and $P_{12}$ eigenmodes of the ring resonator. (c) Measured and (d) calculated $|S_{11}|^2$ parameter for the full structure (ring resonator coupled to the bus waveguide; red solid lines) and for the bus waveguide alone (gray dashed lines), excited by a WR137 waveguide. The data are normalized to the maximum value obtained for the structure with the ring. (e) Measured and (f) calculated absolute square of the difference between $S_{11}$ parameters with and without ring for several values of the gap between the ring and the bus waveguide. The curves are normalized to the maximum value corresponding to the smallest gap.}
	\label{fig:microwave_1}
\end{figure*}

\section{Excitation of the ring resonances via a bus waveguide}

The ring resonances can be excited in a conventional manner using a single-mode side-coupled bus waveguide, see the inset in Fig.~\ref{fig:optical_2}(a). Figure~\ref{fig:optical_2}(a) shows a representative transmission spectrum for a bus waveguide of width 160~nm and a gap of 320~nm between the waveguide and the ring, which is close to the critical-coupling condition. To relate the observed resonances to the ring eigenmodes in Fig.~\ref{fig:optical_1}(a), we note that modes belonging to the lower dispersion branch [Fig.~\ref{fig:optical_1}(f)] exhibit suppressed fields on the outer side of the resonator and enhanced fields on the inner side, whereas the opposite holds for modes from the upper branch [Fig.~\ref{fig:optical_1}(g)]. This near-field interference is analogous to the Kerker effect for waveguide modes~\cite{SavelevJAP2019} and has been exploited to realize flat bands and topologically nontrivial waveguide arrays~\cite{MikhinNL2023}. Owing to this asymmetric field distribution, excitation of the ``P'' modes is significantly less efficient for the chosen gap: as shown in Fig.~\ref{fig:optical_2}(b), excitation of the ``N'' modes with $m=24,25,26$ reduces the transmission to nearly zero, whereas the ``P'' modes are excited with substantially lower efficiency.
 
Note that, in the considered configuration, coupling via the bus waveguide mixes the nearly degenerate ring eigenmodes with opposite azimuthal numbers $\pm m$ only weakly and therefore has a negligible impact on the polarization properties of the excited fields. The resulting frequency splitting $\delta\omega$ between the hybridized even and odd modes (with respect to the $y=0$ mirror plane) is approximately an order of magnitude smaller than the resonance linewidth $\Delta\omega_r$ (full width at half maximum) in the critical coupling regime, i. e. $\delta\omega \ll \Delta\omega_r$. This is also evident from the spatial distributions of the spin density $\tilde{s}_z^{(E)}$ shown in Figs.~\ref{fig:optical_2}(b,c), evaluated at the transmission dips corresponding to the P$_{24}$ and N$_{26}$ resonances. When sampled at angular increments of $2\pi/N$, the envelopes of $\tilde{s}_z^{(E)}$ remain nearly constant, indicating only a minimal admixture of the P$_{-24}$ and N$_{-26}$ modes. Consequently, the DoCP averaged over the area above the resonator, $|\langle C\rangle|\approx 0.2$ and coincides with that obtained for the eigenmode field distribution. As further corroborated by the coupled mode theory analysis in the Supporting Information, Sec.~S2, the average DoCP is relatively robust to this splitting and decreases by only $\sim 30\%$ when $\delta\omega$ becomes comparable to $\Delta\omega_r$.

\section{Experimental results for the microwave spectral range}

Next, we experimentally investigate the proposed resonator in microwave spectral range. To this end, we fabricate a ring resonator consisting of 40 coaxially arranged ceramic cylinders with height $h=5~\mathrm{mm}$, radius $r=10~\mathrm{mm}$, relative permittivity $\varepsilon \approx 10$, and loss tangent $\tan\delta \approx 2\times 10^{-4}$. The ring diameter, measured between the cylinder centers, is $D_{\mathrm{ring}}=127~\mathrm{mm}$. For experimental simplicity, the bus waveguide is implemented as a straight chain of identical cylinders with a period of~$\approx 18~\mathrm{mm}$. Excitation and reflection measurements are performed using a commercial WR137 waveguide positioned on one side of the bus waveguide at a distance of~$\approx 4~\mathrm{mm}$; the corresponding spectral response is characterized via the $S_{11}$ parameter. In addition, the electric field above the resonator is probed with an open-ended coaxial-cable antenna. A schematic of the experimental setup is shown in Fig.~\ref{fig:microwave_1}(a), and a photograph is provided in Fig.~S3 of the Supporting Information.

Similar to the optical design, the isolated ring resonator supports $2N=80$ eigenmodes (see Fig.~S4 in the Supporting Information). Figs.~\ref{fig:microwave_1}(b) show two representative modes near $6.5~\mathrm{GHz}$. The ``N''-type mode with azimuthal number $m=14$ exhibits a field profile enhanced on the outer side of the ring and quasi-uniform $z$-component of the optical spin density in the plane $z=r+2~\mathrm{mm}$. When averaged over this plane, the DoCP reaches $\langle C \rangle \approx -0.71$. In contrast, for the ``P''-type mode with $m=12$ the fields are strongly suppressed on the outer side, while $s_z^{(E)}$ is predominantly positive, yielding $\langle C \rangle \approx +0.6$.

Figures~\ref{fig:microwave_1}(c,d) show the measured and calculated $|S_{11}|^{2}$ reflection spectra, respectively, for the bus waveguide alone and for the bus waveguide coupled to the ring resonator. The oscillatory background observed for the bus waveguide originates from a sequence of Fabry-P$\mathrm{\acute{e}}$rot-like resonances formed by relatively strong partial reflections at the waveguide end [gray dashed curves in Figs.~\ref{fig:microwave_1}(c,d)]. Note that, although the bus waveguide supports two modes---symmetric and antisymmetric with respect to the $x=0$ mirror plane---the symmetric placement of the WR137 waveguide results in the excitation of only one mode type. When the ring is introduced, additional resonances appear on top of the background spectrum (red solid curves in Fig.~\ref{fig:microwave_1}(c,d)). The numerical and experimental results are in good agreement. The observed resonance frequencies also coincide with those of the ``N''-type eigenmodes of the isolated ring (shown only in Fig.~S4 of the Supporting Information). In contrast, the ``P''-type resonances are not visible in the reflection spectra because their fields are strongly suppressed on the outer side of the ring, as illustrated in the upper-right panel of Fig.~\ref{fig:microwave_1}(b). To quantify the contribution of ring resonances with opposite azimuthal numbers $\pm m$ to the measured response, Figs.~\ref{fig:microwave_1}(e,f) show the measured and calculated squared difference between the reflection spectra with and without the ring resonator, focusing on the $N_{14}$ resonance near $6.5~\mathrm{GHz}$. For a large gap $g_c$ between the ring and the bus waveguide, only a single peak is observed, indicating that the spectral splitting between the even and odd modes is much smaller than their linewidths. As a result, the two resonances overlap almost completely, and the response is dominated by the mode with positive $m$ (solid black curve in Figs.~\ref{fig:microwave_1}(e,f)). Reducing the gap increases the coupling between the ring and bus waveguide modes and leads to a noticeable splitting between the even and odd modes. The Fano-like line shape arises from reflections at the bus waveguide ends, as discussed in more detail in Sec.~S2 of the Supporting Information within the coupled mode theory framework.

\begin{figure}[h]
\centering
\includegraphics[width=1\columnwidth]{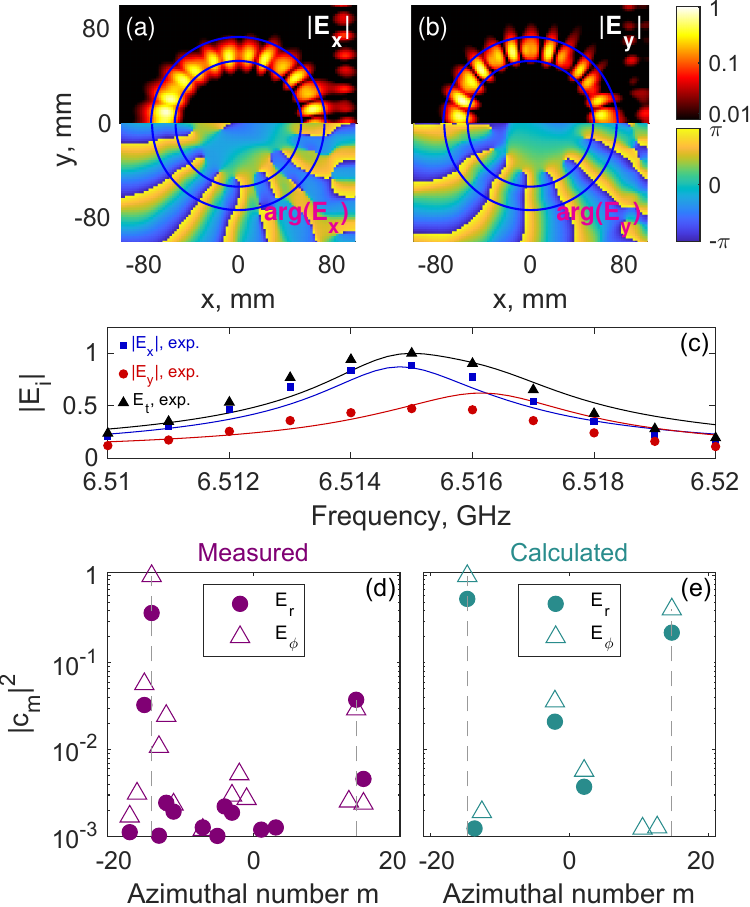}
	\caption{(a,b) Measured amplitudes (top) and phases (bottom) of the (a) $E_x$ and (b) $E_y$ components at a height of $\approx 2~\mathrm{mm}$ above the ring, at $6.515~\mathrm{GHz}$.(c) Simulated (solid lines) and measured (markers) spectra of $|E_x|$, $|E_y|$, and $E_t=\sqrt{|E_x|^2+|E_y|^2}$, normalized to the maximum of $E_t$. (d,e) Angular power spectrum $|c_m|^2$ as a function of azimuthal number $m$ obtained from the (d) measured and (e) calculated $E_r(\phi)$ and $E_{\phi}({\phi})$ distributions.}
	\label{fig:microwave_2}
\end{figure}

The measured amplitude and phase distributions of the $E_{x}$ and $E_{y}$ field components, recorded at a height of $\approx 2~\mathrm{mm}$ above the ring, are shown in Figs.~\ref{fig:microwave_2}(a,b) for a gap of $\approx 7.5~\mathrm{mm}$. The magnitudes $|E_x|$ and $|E_y|$ are of the same order, while the phase pattern confirms that the resonance corresponds to the azimuthal number $m=14$. To estimate the even--odd mode splitting in this configuration, Fig.~\ref{fig:microwave_2}(c) presents the spectra of $|E_x|$, $|E_y|$, and $E_t=\sqrt{|E_x|^2+|E_y|^2}$, normalized to the maximum of $E_t$ at the point $(x,y)=(-R_{\mathrm{ring}},0)$. The measured data (markers) agree well with the simulations (solid lines), indicating a splitting below $2~\mathrm{MHz}$.

We have further measured the $E_r(\phi)$ and $E_{\phi}(\phi)$ dependence at the same height and for radial coordinate $r=R_{\mathrm{ring}}$. The direct comparison between measured and calculated fields is provided in Supporting Information, Fig.~S5. In Figs.~\ref{fig:microwave_2}(d,e) we plot the square magnitudes $|c_m|^2$ of the harmonics with azimuthal number $m$, calculated as follows:
\begin{equation}
c_m = \dfrac{1}{N} \sum_{n=0}^{N-1}E_n\exp(-im\phi_n),
\end{equation}
where $N=40$, $\phi_n=2\pi n/N$, $E_n$ denotes either $E_r$ or $E_{\phi}$ component sampled above the centres of the cylinders. The coefficients obtained $c_m$ are normalized by the maximum value of $|c_m|$ for both field components.

One can observe in Figs.~\ref{fig:microwave_2}(d,e) that the dominant harmonic has $m=-14$. Notably, the difference between the phases of $c_{-14}$ components of $E_r$ and $E_{\phi}$ is $-0.45\pi$ for measured fields and $-0.55\pi$ for calculated fields. The harmonic with $m=+14$ also provides a noticeable contribution, which is expected due to rather strong reflection from the end of the bus waveguide. For $c_{14}$ component the corresponding phase difference is $+0.71\pi$ for measured fields and $+0.55\pi$. The change of direction of electric field rotation is in accordance with the reverse of the propagation direction. The certain discrepancy between the measured and simulated field components can be attributed to the finite size of the near-field probe. This effect is particularly important for $E_{\phi}$, because both the amplitude and the phase vary significantly along the probe orientation (i. e., with azimuthal angle $\phi$).

\section{Discussion and summary}

To summarize, we have developed a planar integrated optical ring resonator that supports resonances with a quasi-uniform distribution of the optical spin-density component perpendicular to the substrate. This feature makes the proposed structure promising for integrated optical elements based on spin-dependent coupling with quasi–two-dimensional materials, such as single-photon emitters~\cite{MaOptica2022}, non-reciprocal elements~\cite{PakniyatJAP2024}, and optical isolators~\cite{GuddalaNC2021,dushaq2025non}.

Coupling between right- and left-handed circularly polarized dipole sources and an integrated optical waveguide is asymmetric in regions where the optical mode exhibits a non-zero transverse (normal to the monolayer) component of the optical spin. In the case of two-dimensional materials with distributed excitons, this asymmetry requires breaking the vertical mirror symmetry of the waveguide, as it vanishes for a symmetric structure. Consequently, spin-dependent excitation and luminescence reported in the literature rely on shifting the pumping spot away from the symmetry plane~\cite{GongSC2018,GongNL2020,WooOE2021,ChenACSnano2021,ShreinerNP2022}, asymmetric placement of the monolayer flake~\cite{GuddalaNC2021}, positioning of localized emission sites and defects within monolayers~\cite{YangOE2019,MaOptica2022}, and/or asymmetry induced by strain~\cite{dushaq2025non}. These approaches are technologically challenging and can substantially limit device-to-device reproducibility.

In contrast, the non-zero average optical spin density inherent to the design proposed in our work enables strong coupling asymmetry and non-reciprocal phase shifts even with uniform placement of the two-dimensional material. Moreover, the resonant nature of the structure further enhances these effects, potentially allowing the realization of magneto-optical isolation or strong exciton–photon coupling regimes. Finally, we note that the current design can be scaled up or down to operate in the desired spectral range relevant to a particular material platform.

\section{Acknowledgments}
\begin{acknowledgments}
The work was supported by the Russian Science Foundation, project \#25-72-10103, \url{https://rscf.ru/project/25-72-10103/}
\end{acknowledgments}

\bibliographystyle{unsrt}
\bibliography{refs.bib}


\beginsupplement

\section*{Supproting Information: Emergence of transverse optical spin in a subwavelength grating ring resonator}
\section*{Contents}
S1. Coupling of waveguide modes and formation of non-zero total optical spin \dotfill \pageref{sec:SI_spin}\\
S2. Coupled-mode theory \dotfill \pageref{sec:SI_cmt}\\
S3. Details of the microwave design and experiment \dotfill \pageref{sec:SI_mw}\\
\clearpage


\section{S1. Coupling of waveguide modes and formation of non-zero total optical spin}
\label{sec:SI_spin}

\begin{figure*}[h]
\centering
\includegraphics[width=0.9\textwidth]{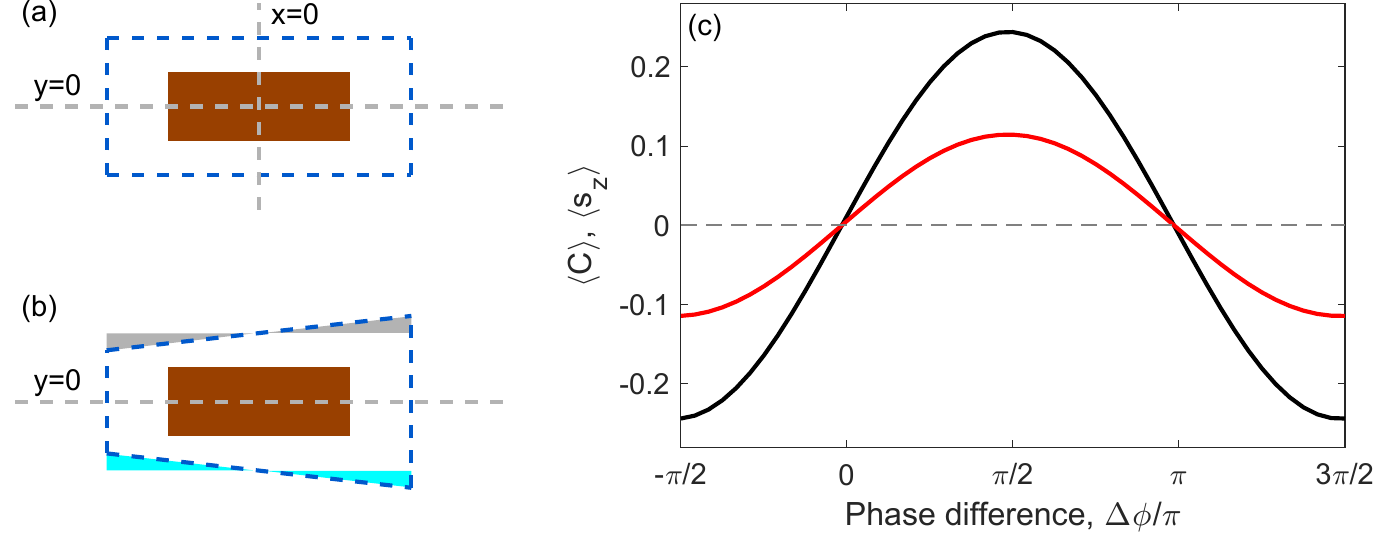}
	\caption{(a,b) Top view of the unit cell of the (a) straight periodic waveguide and (b) unit sector of the ring or bent waveguide; horizontal dashed blue lines indicate the boundaries of the unit cell; gray areas in (b) indicate the perturbation areas. (c) Black curve: degree of circular polarization of electric field averaged over the plane above the waveguide $\langle C \rangle$, [Eq.~\eqref{eqSI:DoCP_av}], calculated for the superposition of the two waveguide modes as a function of their phase difference. Red curve: $z$ component of the spin density integrated over the volume of the whole unit cell and normalized by the corresponding energy density, [Eq.~\eqref{eqSI:Total_spin}].}
	\label{figSI_spin_DoCP}
\end{figure*}

The coupling of the two quasi-degenerate modes induced by waveguide bending, and the resulting formation of a nonzero total $z$-component of the optical spin, are illustrated in Fig.~\ref{figSI_spin_DoCP}. The unit cell of the straight waveguide is shown in Fig.~\ref{figSI_spin_DoCP}(a). The waveguide axis is oriented along the $y$ (vertical) direction. The left and right boundaries are shown schematically, while the actual unit cell is infinite along the $x$ (horizontal) direction. The two waveguide modes exhibit different even/odd symmetries with respect to both symmetry planes, $x=0$ and $y=0$.

The bent waveguide can be viewed as the same structure with a slightly modified unit cell, as illustrated in Fig.~\ref{figSI_spin_DoCP}(b). Breaking the $x=0$ mirror symmetry enables coupling between the two modes, which (within the framework of the perturbation theory) is governed by overlap integrals evaluated over the dashed gray and cyan regions shown in Fig.~\ref{figSI_spin_DoCP}(b). Owing to the opposite spatial symmetry of the modes with respect to the $x=0$ plane and time-reversal symmetry, the overlap integral over the gray region equals $-I^*$, where $I$ denotes the overlap integral over the cyan region. Consequently, when the $y=0$ symmetry is preserved, the total overlap integral is purely imaginary. Then, the fields of the bent waveguide can be represented by the sum of the two modes of the straight waveguide with a relative phase difference of $\pm \pi/2$ between the modes.

The polarization properties of the combined waveguide modes as a function of their relative phase difference are illustrated in Fig.~\ref{figSI_spin_DoCP}(c). The two modes are assumed to have equal amplitudes and are both normalized to unit power. The average degree of circular polarization (DoCP) is calculated following the procedure described in the main text:
\begin{equation}
\langle C \rangle= \dfrac{\langle |E_r|^2 - |E_l|^2 \rangle}{\langle |E_r|^2 + |E_l|^2 \rangle},
\label{eqSI:DoCP_av}
\end{equation}
where $E_{r,l}(\VB{r}) = \VB{E}(\VB{r})\cdot (\hat{x} \mp i\hat{y})$, $\VB{E} = \VB{E}_1 + \exp(i\phi)\VB{E}_2$, $\VB{E}_{1,2}(\VB{r})$ are the complex electric-field amplitude distributions of the first and second waveguide modes, respectively. 
The averaging is performed over a plane normal to the $z$-direction, located directly above the top oxide layer, i.e., 30~nm above the silicon particles. The average energy normalized $z$ component of the optical spin density is calculated as follows:
\begin{equation}
\langle s_z \rangle = \omega \dfrac{\int (s_z^{(E)}(\VB{r}) + s_z^{(H)}(\VB{r})) dV}{\int W(\VB{r}) dV},
\label{eqSI:Total_spin}
\end{equation}
where~\cite{PicardiOPT2018}
\begin{equation}
\VB{s}^{(E)}(\VB{r}) = \dfrac{\eps_0 \eps(\VB{r})}{4 \omega}\Im(\VB{E}^*(\VB{r})\times\VB{E}(\VB{r})), \;\;
\VB{s}^{(H)}(\VB{r}) = \dfrac{\mu_0}{4\omega}\Im(\VB{H}^*(\VB{r})\times\VB{H}(\VB{r})),
\end{equation}
where $W$ denotes the energy density of the electromagnetic field. In this case, the averaging is performed over the entire unit cell. The results of the spin calculations show that, as soon as $\phi \neq 0,\pi$, a nonzero $\langle s_z \rangle$ emerges. The electric and magnetic contributions to the optical spin are nearly equal (not shown in the figure). The maximum spin magnitude is reached for $\phi = \pm \pi/2$, which is realized in the ring resonator. Calculations of $\langle C \rangle$ further indicate that the average degree of circular polarization exceeds $0.2$ for $\phi = \pm \pi/2$, which can be employed to realize the average asymmetric coupling to valley-polarized excitons in two-dimensional materials placed above the resonator.

\newpage
\section{S2. Coupled-mode theory}
\label{sec:SI_cmt}

\begin{figure*}[h]
\centering
\includegraphics[width=0.7\textwidth]{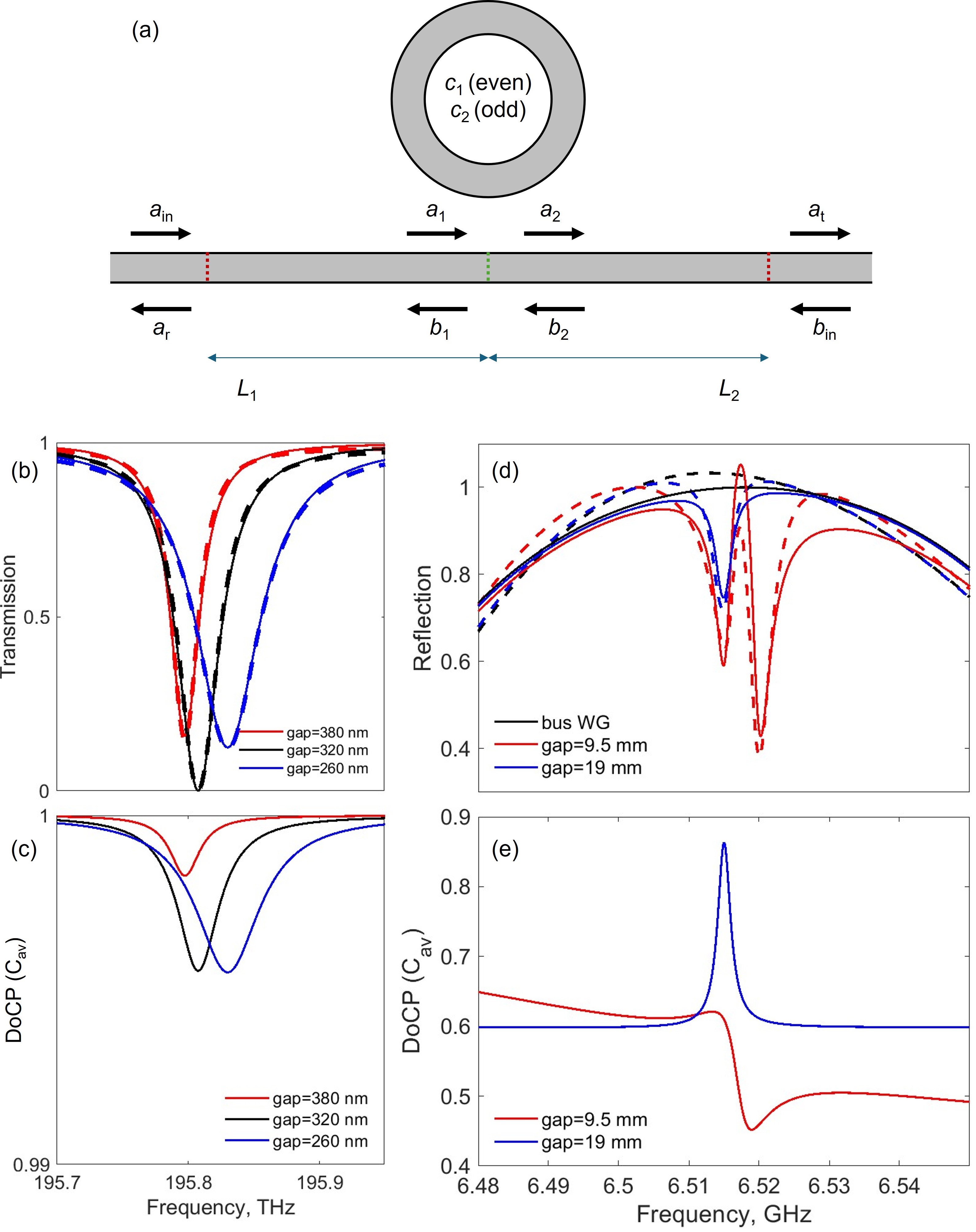}
	\caption{(a) Scheme of the model employed in the analysis of the spectral response of the studied structures: bus waveguide with partially reflecting elements (red dots) is coupled to the ring resonator. (b,c) Optical design: (b) Transmission $T_{tot}$ for three values of the gap between the waveguide and the resonator; dashed lines correspond to numerical simulations, solid lines -- to the CMT results. (c) Degree of circular polarization calculated with Eq.~\eqref{eq:CMT_Cav} for the same parameters as in (b). (d,e) Microwave design: (d) Reflection $R_{tot}$ for bus waveguide only (black lines) and from bus+ring for gap 19~mm (blue lines) and 9.5~mm (red lines); dashed lines correspond to numerical simulations, solid lines -- to the CMT results. (c) Degree of circular polarization calculated with Eq.~\eqref{eq:CMT_Cav} for the same parameters as in (e).  }
	\label{figSI_CMT}
\end{figure*}
The calculated and measured spectral characteristics of the considered ring+bus waveguide systems can be interpreted within the framework of coupled mode theory. In order to analyze both optical and microwave designs employed in the main text, first, we write down the general transfer matrix for the system, schematically shown in Fig.~\ref{figSI_CMT}(a), consisted of a single-mode waveguide with a pair of reflectors and a side-coupled cavity between them supporting two modes, one even (index 1) and one odd (index 2)~\cite{FanAPL2002,SuhIEEE2004}:
\begin{equation}
\begin{pmatrix}
a_t \\
b_{in}
\end{pmatrix}
=
\VB{M}_{tot}
\begin{pmatrix}
a_{in} \\
a_r
\end{pmatrix}
=
\VB{M}_{out} \, \VB{M} \, \VB{M}_{in}
\begin{pmatrix}
a_{in} \\
a_r
\end{pmatrix}
\end{equation}
where $\VB{M}_{in}$ ($\VB{M}_{out}$) is the transfer matrix through the first (second) reflector in general case reads:
\[
\begin{pmatrix}
a_1 \\
b_1
\end{pmatrix}
=
\VB{M}_{in}
\begin{pmatrix}
a_{in} \\
a_{r}
\end{pmatrix}
=
\frac{1}{|t|}
\begin{pmatrix}
e^{i(\theta+\varphi_1)} & |r| e^{i(\delta_1 - \theta + \varphi_1)} \\
|r| e^{-i(\delta_1 - \theta + \varphi_1)} & e^{-i(\theta+\varphi_1)}
\end{pmatrix}
\begin{pmatrix}
a_{in} \\
a_{r}
\end{pmatrix},
\]

\[
\begin{pmatrix}
a_t \\
b_{in}
\end{pmatrix}
=
\VB{M}_{out}
\begin{pmatrix}
a_{2} \\
b_{2}
\end{pmatrix}
=
\frac{1}{|t|}
\begin{pmatrix}
e^{i\theta+\varphi_2} & -|r| e^{-i(\delta_1 - \theta + \varphi_2)} \\
-|r| e^{i(\delta_1 - \theta + \varphi_2)} & e^{-i(\theta+\varphi_2)}
\end{pmatrix}
\begin{pmatrix}
a_{2} \\
b_{2}
\end{pmatrix},
\]
where $t = |t|\exp(i\theta)$ is the transmission coefficients through a reflecting element, $r_1 = |r|\exp(i\delta_1)$, $r_2 = |r|\exp(i\delta_2)$ are reflection coefficients from the inside and outside, respectively, and relation $2\theta = \delta_1 + \delta_2 + \pi$ due to unitarity of the scattering matrix is employed.

The transfer matrix $\VB{M}$ for the ring resonator is expressed through the corresponding scattering matrix $\VB{S}$:
\[
\begin{pmatrix}
b_1 \\
a_2
\end{pmatrix}
=
\VB{S}
\begin{pmatrix}
a_1 \\
b_2
\end{pmatrix}
=
\begin{pmatrix}
-\dfrac{\gamma_1}{\Delta_1} + \dfrac{\gamma_2}{\Delta_2}
&
1 - \dfrac{\gamma_1}{\Delta_1} - \dfrac{\gamma_2}{\Delta_2}
\\[1em]
1 - \dfrac{\gamma_1}{\Delta_1} - \dfrac{\gamma_2}{\Delta_2}
&
-\dfrac{\gamma_1}{\Delta_1} + \dfrac{\gamma_2}{\Delta_2}
\end{pmatrix}
\begin{pmatrix}
a_1 \\
b_2
\end{pmatrix},
\]
\[
\begin{pmatrix}
a_2 \\
b_2
\end{pmatrix}
=
\VB{M}
\begin{pmatrix}
a_1 \\
b_1
\end{pmatrix}
=
\frac{1}{S_{12}}
\begin{pmatrix}
-|\VB{S}| & S_{22} \\
- S_{11} & 1
\end{pmatrix}
\begin{pmatrix}
a_1 \\
b_1
\end{pmatrix},
\]
where $\Delta_j = i(\omega - \omega_j) + \gamma_j + \gamma^{in}_j$, $\omega_j$ is the resonance frequency of the $j$th resonance, $\gamma_j$ is the coupling coefficient of the $j$th mode to the bus waveguide, $\gamma^{in}_j$ denotes internal losses of the $j$th mode, including radiative and non-radiative losses, and $|\VB{S}|$ is the determinant of $\VB{S}$.
Total transmission and reflection coefficients for excitation from the left ($b_{in}=0$) are expressed via matrix elements as follows:
\[
T_{tot} = |t_{tot}|^2, \,\, t_{tot} = \dfrac{|\VB{M}_{tot}|}{M_{tot_{22}}},\\
R_{tot} = |r_{tot}|^2, \,\, r_{tot} = -\dfrac{M_{tot_{21}}}{M_{tot_{22}}}.
\]
The amplitude of the even and odd modes can be calculated as follows:
\[
\begin{pmatrix}
c_1 \\
c_2
\end{pmatrix}
=
\begin{pmatrix}
i\dfrac{\sqrt{\gamma_1}}{\Delta_1}(a_1+b_2) \\
\dfrac{\sqrt{\gamma_2}}{\Delta_2}(-a_1+b_2)
\end{pmatrix}.
\]

The degree of circular polarization is expressed via the amplitude of the modes assumin equal polarization weights as follows:
\begin{equation}
C = -2\dfrac{\mathrm{Im}(c_1c_2^*)}{|c_1|^2 + |c_2|^2}.
\label{eq:CMT_Cav}
\end{equation}

\textbf{Optical design:} In the optical design (Fig.~3 in the main text) the bus waveguide is assumed to be infinite and, therefore, reflection coefficients are zero $r_1=r_2=0$. Further assuming $\gamma_1=\gamma_2 = \gamma$, $\gamma^{in}_{1}=\gamma^{in}_{2} = \gamma^{in}$ one can write down simple analytical expressions for the total reflection coefficient, total transmission coefficient and DoCP:
\[
R_{tot} = |S_{11}|^2 = \gamma^2\left| \dfrac{1}{\Delta_2} - \dfrac{1}{\Delta_1} \right|^2,\\
T_{tot} = |S_{21}|^2 = \left| 1 - \dfrac{\gamma}{\Delta_2} - \dfrac{\gamma}{\Delta_1} \right|^2,
C_{av} =  2\dfrac{\mathrm{Re}(\Delta_1^*\Delta_2)}{|\Delta_1|^2 + |\Delta_2|^2},
\]
In case of zero coupling between CW and CCW modes, $\omega_1=\omega_2$, $\Delta_1 = \Delta_2$, $R_{tot}=0$, $C=1$, i.e. only a mode with a particular sign of azimuthal number $m$ is excited. In case of non-zero coupling we can define the resonance frequencies of the even and odd modes as $\omega_{1,2} = \omega_0 \mp \delta\omega/2$. Due to non-zero $\delta\omega$ reflection becomes non-zero and DoCP deviates from unity. To quantify this effect we consider the critical coupling regime, i.e. when $T_{tot}$ goes to zero at frequency $\omega=\omega_0$, which is achieved for $\gamma^2 = \gamma_{in}^2 + \delta\omega^2/4$. Then at $\omega=\omega_0$ frequency $R_{tot}$ exhibits maximum and $C_{av}$ exhibits minimum:
\[
R_{tot} = \dfrac{\delta\omega^2}{4(\gamma+\gamma_{in})^2},\\ C_{av} = \dfrac{(\gamma+\gamma_{in})^2 - \delta\omega^2/4}{(\gamma+\gamma_{in})^2 + \delta\omega^2/4}.
\]
The minimal reflection of the order of $R_{tot}\approx 0.01$ observed in numerical simulations implies that the splitting between the CW and CCW modes is negligible for the considered parameters, i.e. $\delta_\omega \ll \gamma_{in}$ (which is also confirmed by direct numerical calculation of $\delta\omega$). Consequently, the average DoCP is very slightly affected by the presence of the bus waveguide, as shown in Figs.~\ref{figSI_CMT}(b,c). The parameters of the structure are the same as in the main text: refractive indices of silicon and silicon oxide $n=3.5$, $n_{\mathrm{SiO_2}}=1.45$; dimensions of the ring $w_x=820$~nm, $w_z=440$~nm, $w_y=170$~nm, $R_{ring}=300$~nm, $N=64$, width of the bus waveguide $w_{bus}=160$~nm. Parameters of the CMT model: for the gap 260~nm: $\omega_0=195.83$~THz, $\delta\omega=2.8$~GHz, $\gamma=20$~GHz; for the gap 320~nm: $\omega_0=195.8075$~THz, $\delta\omega=1.8$~GHz, $\gamma=9.6$~GHz; for the gap 380~nm: $\omega_0=195.7974$~THz, $\delta\omega=0.8$~GHz, $\gamma=4.1$~GHz; $\gamma^{in} = 9.5$~GHz for all gaps. Even for splitting as large as $\delta\omega = 2\gamma^{in}$, $C_{av}$ drops only to $\approx$ 70\%.

\textbf{Microwave design:} In the microwave experiment the ends of the utilized periodic bus waveguide quite strongly reflect the propagating modes. Moreover, the size of the bus waveguide is larger than in the optical design, i.e. perturbation caused by the bus waveguide is stronger. This leads to (i) Fabry-P$\mathrm{\acute{e}}$rot (FP) like oscillations in the bus waveguide appearing in the background spectra, (ii) larger splitting between CW and CCW modes, (iii) appearance of Fano-type lineshapes due to interference between FP and ring resonances.

To reproduce the measured spectra we choose the following parameters of the model. Phases $\phi_1,\phi_2$, acquired during propagation of the waveguide are chosen based on the dispersion and the length of the bus waveguide. Phase $\theta=\pi/2$ is chosen to satisfy the FP resonance condition. Reflection coefficient $|r|=0.5$ is chosen to mimic the reflection spectrum of the bus waveguide, i.e. to have the similar width of the FP resonances, while $|t| = \sqrt{1-|r|^2}$. The resonance frequencies are chosen as $\omega_{1,2} = \omega_r \mp \delta\omega/2$, where central frequency $\omega_r$ and splitting $\delta\omega$ are determined from the eigenmode simulations. Internal losses are determined from the eigenmodes simulations $\gamma^{in}_1=\gamma^{in}_2=1\cdot10^{-3}$~GHz. Further, by adjusting the values of $\delta_1$, $\gamma_1$, $\gamma_2$, we reconstruct numerically simulated reflection spectra, as shown in Fig.~\ref{figSI_CMT}(d). Fig.~\ref{figSI_CMT}(e) indicates that the DoCP decreases due to reflection from the end of the waveguide, however, can still be close to 1 at resonance for the weak coupling between the waveguide and the resonator. We believe that the correspondence between the analytical model and simulations can be further improved if one takes into account (i) the coupling between the WR137 waveguide and the bus waveguide, (ii) the presence of the second (antisymmetric) mode in the bus waveguide, which is not directly excited by the WR137 waveguide, but contributes to the coupling with the ring resonator.

\newpage
\section{S3. Details of the microwave design and experiment}
\label{sec:SI_mw}

\begin{figure*}[h]
\centering
\includegraphics[width=0.55\textwidth]{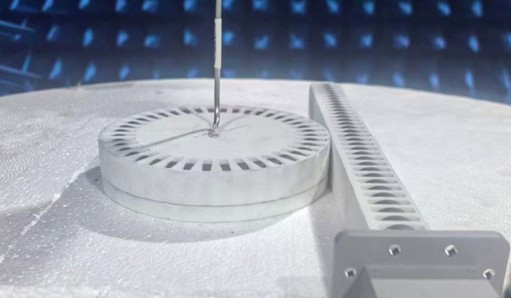}
	\caption{A photograph of the experimental setup: A ring resonator composed of ceramic cylinders is excited through the bus waveguide composed of the same cylinders. The bus waveguide is excited via symmetrically-placed WR137 commercial waveguide. Bus waveguide and the ring are placed in a foam holder with permittivity $\approx 1$. The fields above the resonator are measured with an open-ended coaxial cable.}
	\label{fig:SI_photo}
\end{figure*}

\begin{figure*}[h]
\centering
\includegraphics[width=1\textwidth]{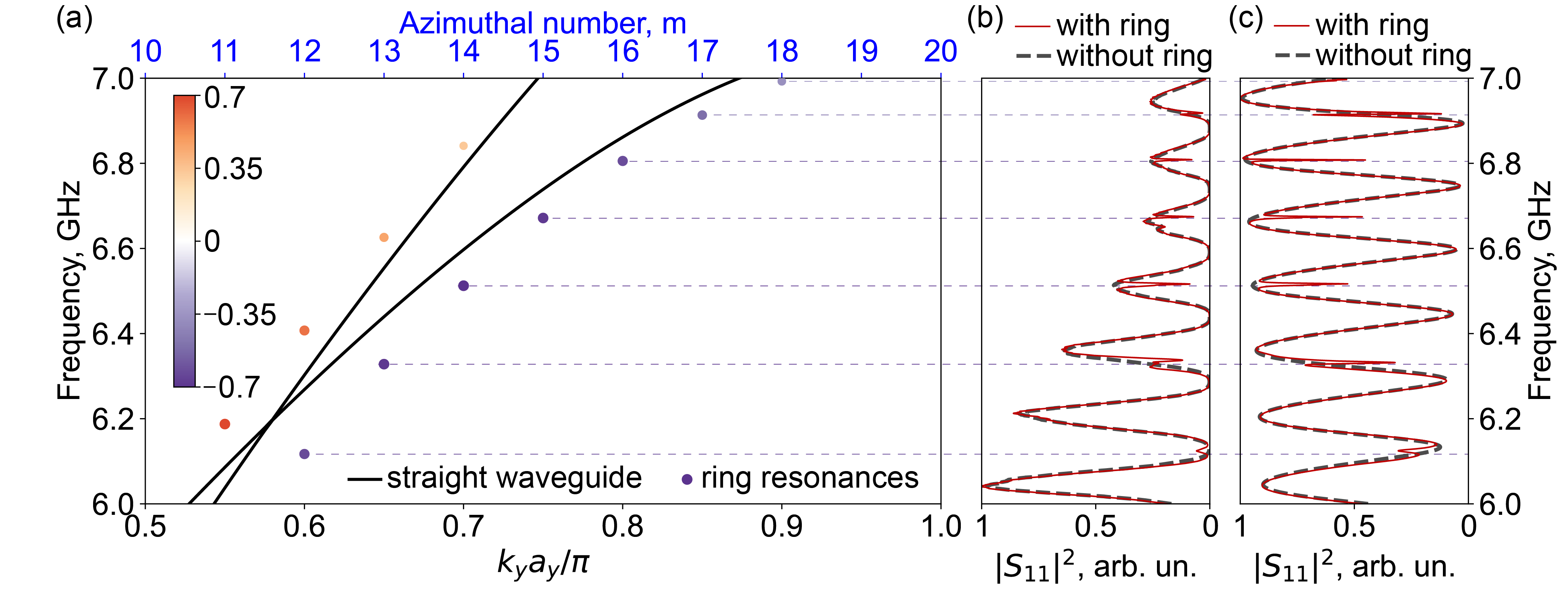}
	\caption{Characteristics of the microwave ring resonator. (a) Dispersion of a straight SWG (solid lines) and resonance frequencies of the ring resonator based on such a waveguide; purple dots correspond to resonances with negative average $s_z^{(E)}$, orange -- with positive $s_z^{(E)}$; size of the circles and  their colour reflect the value and the sign of the average $s_z^{(E)}$; dashed lines indicate the resonances that are excited in experiment via the bus waveguide. (b) Measured and (c) calculated $S_{11}$ parameter for the bus waveguide only (dashed gray curves) and bus waveguide with ring resonator (red solid lines) under excitation with a WR137 waveguide. See parameters of the ring--bus waveguide configuration in main text.}
	\label{fig:SI_disper}
\end{figure*}

\begin{figure}[h]
\centering
 \includegraphics[width=0.75\columnwidth]{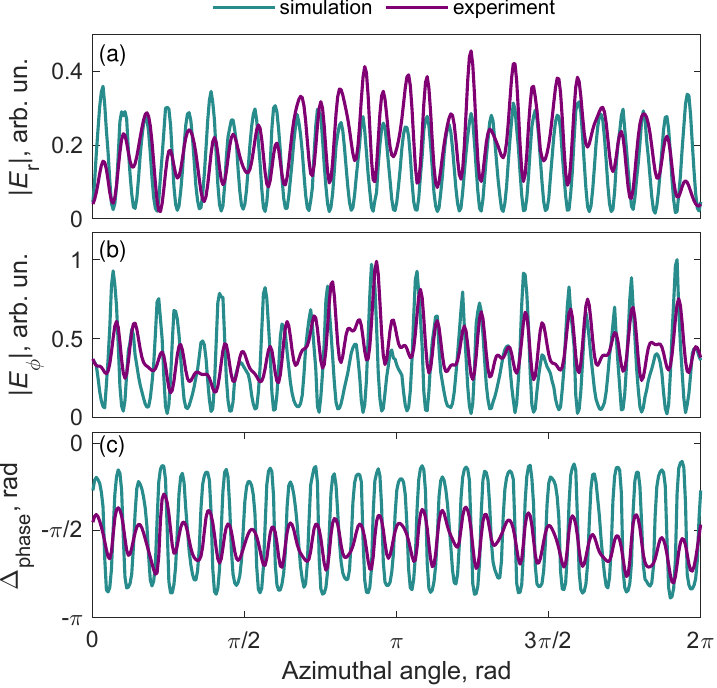}
	\caption{(a,b) Measured and calculated amplitudes of (a) $|E_r(\phi)|$ and (b) $|E_{\phi}(\phi)|$ electric field components. (c) Phase difference of $E_r$ and $E_{\phi}$. Fields are taken at radial coordinate $r=R_{ring}$ and $z$ coordinate $\approx2$~mm above the ceramic cylinders.}
	\label{figSI:fields}
\end{figure}


Figure~\ref{figSI:fields}(a-c) compares the measured and calculated $E_r$ and $E_{\phi}$ components, as well as their relative phase, along the ring circumference. The observed amplitude oscillations mainly appear due to interference between the counter-propagating modes with azimuthal numbers $\pm m$, caused by reflection from the end of the bus waveguide. The phase difference exhibits similar oscillations around the mean value of $-\pi/2$, indicating the quasi-circular polarization of the electric field in the $xy$ plane. In the experiment, these oscillations are further reduced due to averaging of the fields over the finite probe size.


\end{document}